\documentclass[aps,amsmath,amssymb, preprint, 14pt]{revtex4-1}
\usepackage{amsmath, latexsym, color, graphicx, tabularx, booktabs, enumerate}
\usepackage{mathptmx}
\usepackage{multirow}
\usepackage{dcolumn}
\usepackage{bm}
\usepackage{cleveref}
\usepackage{epstopdf}
\date{\today}

\usepackage[normalem]{ulem} 

\begin{document}
\author{Adrienn Ruzsinszky$^{1}$}
\author{Niraj K. Nepal$^{1}$}
\author{J.M. Pitarke$^{2,3}$}
\author{John P. Perdew$^{1,4}$}

\affiliation{$^1$Department of Physics, Temple University, Philadelphia, PA USA 19122}
\affiliation{$^2$CIC nanoGUNE BRTA and DIPC, Donostia, Basque Country, Spain}
\affiliation{$^3$Materia Kondentsatuaren Fisika Salida and Centro Fisica Materiales CSIC-UPV/EHU, Bilbao, Basque Country, Spain}
\affiliation{$^4$Department of Chemistry, Temple University, Philadelphia, PA USA 19122}

\title{Constraint-based Wavevector- and Frequency-dependent Exchange-Correlation Kernel of the Uniform Electron Gas}
\begin{abstract}
\noindent According to time-dependent density functional theory, the exact exchange-correlation kernel f$_{xc}$(n, q, $\omega$) determines not only the ground-state energy but also the excited-state energies/lifetimes and time-dependent linear density response of an electron gas of uniform density n $=$ 3/(4$\pi$r$^3_s$). Here we propose a parametrization of this function based upon the satisfaction of exact constraints. For the static ($\omega$ = 0) limit, we modify the model of Constantin and Pitarke at small wavevector q to recover the known second-order gradient expansion, plus other changes. For all frequencies $\omega$ at q $=$ 0, we use the model of Gross, Kohn, and Iwamoto. A Cauchy integral extends this model to complex $\omega$ and implies the standard Kramers-Kronig relations. A scaling relation permits closed forms for not
only the imaginary but also the real part of f$_{xc}$ for real $\omega$. We then combine these ingredients by damping out the $\omega$ dependence at large q in the same way that the q dependence is damped. Away from q $=$ 0 and $\omega$ $=$ 0, the correlation contribution to the kernel becomes dominant over exchange, even at r$_s$ $=$ 4, the valence electron density of metallic sodium. The resulting correlation energy from integration over imaginary $\omega$ is essentially exact. The plasmon pole of the density response function is found by analytic continuation of f$_{xc}$ to $\omega$ just below the real axis, and the resulting plasmon lifetime first decreases from infinity and then
increases as q grows from 0 toward the electron-hole continuum. A static charge-density wave is found for r$_s$ $>$ 69, and shown to be associated with softening of the plasmon mode. The exchange-only version of our static kernel confirms Overhauser's 1968 prediction that correlation enhances the charge-density wave.
\end{abstract}
\maketitle

\section{I\MakeLowercase{ntroduction}}
\noindent In time-dependent density functional theory (TDDFT) \cite{1,2,3}, the exact linear
density response function $\chi$(\textbf{r}, \textbf{r$^\prime$}, $\omega$) of an electronic system in its ground state to
a weak external scalar potential $\delta$v(\textbf{r$^\prime$}, $\omega$), oscillating at angular frequency $\omega$, provides access to the exact ground- and excited-state energies of the system. Under the standard assumption that the ground-state and time-dependent densities for the real interacting system are the same as those of a fictitious non-interacting system in an effective scalar potential (the Kohn-Sham or KS potential in the ground-state case), the true response function $\chi$ can be constructed from the calculable non-interacting response function $\chi_{\text{KS}}$ and an exchange-correlation kernel f$_{xc}$. For a homogeneous system like the uniform electron gas, where  $\chi$ and $\chi_{\text{KS}}$ can only depend upon $\mid$\textbf{r$^\prime$} - \textbf{r}$\mid$, Fourier transformation leads to the simple algebraic solution
\begin{equation}
\chi(q,\omega) = \frac{\chi_{\text{KS}}(q,\omega)}{\epsilon(q, \omega)}
\label{eq1}
\end{equation}

\begin{equation}
\epsilon(q,\omega) = 1 - \left[\frac{4\pi}{q^2} + f_{xc}(q,\omega)\right] \chi_{\text{KS}}(q,\omega).
\label{eq2}
\end{equation}

\noindent The amplitude of the density response to a weak perturbation of amplitude $\delta v$ is $\delta n = \chi \delta v$. $\chi_{\text{KS}}$ is the Lindhard function \cite{4}, and the dependence of all functions upon the uniform density n is implicit. Through the adiabatic-connection fluctuation-dissipation theorem \cite{5,6,7},  $\chi$ yields the ground-state exchange-correlation energy from an integral along the upper half of the imaginary frequency axis, under analytic continuation to complex frequencies. The poles of the response functions in the lower-half complex plane are the excitation energies/inverse lifetimes at wavevector q. The function $\epsilon$ can vanish, introducing a collective excitation or plasmon that is not present in the non-interacting KS system. Many exact properties of the exchange-correlation kernel have been derived, and models have been constructed to satisfy those exact constraints, in much the same way that the density functional for the ground-state exchange-correlation energy, E$_{xc}$[n], is often approximated by the satisfaction of known exact constraints. Note that the uniform-gas correlation energy from quantum Monte Carlo (QMC) calculations \cite{8}, which, as extended and parametrized in Refs. \cite{9,10,11}, typically serves as an input to the construction of such functionals, can be accurately predicted (not fitted) by a constraint-based interpolation \cite{12,13} between known high- and low-density limits.\\

In this work, we will develop a constraint-based model kernel that refines
the Constantin-Pitarke 2007 (CP07) \cite{14} q-dependent static ($\omega$ = 0) kernel, and combines it with the Gross-Kohn-Iwamoto \cite{2,15} dynamic kernel for q $=$ 0. Our
kernel is developed for general complex frequencies, while that of CP07 is
developed only for zero and imaginary frequencies. While many practical calculations with
TDDFT for real systems use the adiabatic local density approximation based upon
the uniform-gas f$_{xc}$(q $=$ 0, $\omega$ $=$ 0), we will see that there are strong dependences on both variables, and that, away from q $=$ 0 and $\omega =$ 0, the correlation contribution can dominate over exchange, even at the valence electron densities of metals. Our model passes several early tests: It yields very accurate correlation energies per electron for the uniform gas without fitting (while the dynamic CP07 is fitted to those energies), predicts finite lifetimes for plasmons \cite{16} of small non-zero wavevector, and finds at about the right low density a static charge-density wave \cite{8,17,18} that arises from a softening of the plasmon mode. We did not find a charge-density wave at any density with the original CP07 kernel. We hope that our model will have other applications, and that it might have implications for TDDFT in real systems.\\

The static kernel f$_{xc}$(q, $\omega =$ 0) for the uniform electron gas (jellium) has been calculated via QMC \cite{19} and parametrized by Corradini et al. \cite{20}. Jellium is an important model because it has a Hamiltonian with Coulomb repulsions between electrons, but with the external potential simplified from that of positive ions to that of a uniform positive-charge background. If this background is allowed to expand or contract, then jellium is only stable for an electron density n $=$ 3/(4$\pi$r$^3_s$) with r$_s$ $\approx$ 4 (in atomic units or bohr), near the valence-electron density of metallic sodium. But, by adding an appropriately-chosen short-range contribution to the external potential, a stabilized jellium model can be constructed for a better description of all simple metals and their surfaces \cite{21}. Since the bulk electron density remains uniform, all exchange-correlation effects in bulk stabilized jellium are the same as those in bulk jellium at the same density. The book by Giuliani and Vignale \cite{22} provides a detailed discussion of exchange and correlation in the uniform electron gas, and an explanation of the important difference between the short-range kernel of the uniform electron gas (where f$_{xc}$(q, $\omega$) tends
to a finite constant as q $\rightarrow$ 0) and the ultra-nonlocal kernels of other systems (where this limit diverges $\sim$ $\frac{1}{q^2}$). For this and other reasons, kernel development has been generalized \cite{23,24,25} from density to current-density functionals.\\

Even before TDDFT, it was known that there is a local field factor G$_{xc}$(q, $\omega$), with
\begin{equation}
f_{xc}(q, \omega) = -\left(\frac{4\pi}{q^2}\right)G_{xc}(q,\omega),
\label{eq3}
\end{equation}
\noindent that corrects the over-estimation of short-range correlation in the random phase approximation (RPA) for the correlation energy of a uniform electron gas. Singwi and collaborators \cite{26} modelled a static local field factor that essentially predicted the uniform-gas correlation energy later found from QMC \cite{8}. Lein, Gross and Perdew \cite{27} used the Richardson-Ashcroft (RA) \cite{28} dynamic kernel (as developed for imaginary frequencies only) to show that the static limit of a good kernel can capture most of the correction to the RPA correlation energy, although including the frequency dependence of the RA kernel gives even more-accurate correlation energies. Here we will find that the static version of our kernel already predicts very accurate correlation energies, which are hardly changed by inclusion of our frequency dependence.

\section{D\MakeLowercase{ensity dependence of constraint-based kernels}}

For simplicity, we will discuss here the density dependence of constraint-
based static kernels for the uniform gas. The frequency dependence
complicates the notation but does not change the conclusions. Here we will use
the Fermi wavevector k$_\text{F}$ $=$ $(3\pi^2 n)^{1/3}$. The non-interacting response function has the simple scaling equality $\chi_{\text{KS}}$(q,0) $=$ k$_\text{F}$F($\frac{q}{2k_\text{F}}$), and of course the Fourier transform of the Coulomb interaction between electrons is $\frac{4\pi}{q^2}$ $=$ k$^{-2}_\text{F}\pi(\frac{q}{2k_\text{F}})^{-2}$.\\

The kernel has Coulomb-like scaling equalities only in the high-density and
low-density limits
\begin{equation}
f_{xc}(q,0) \rightarrow k^{-2}_\text{F} G\left(\frac{q}{2k_\text{F}}\right) \hspace{1in} (k_\text{F} \rightarrow\infty)
\label{eq4}
\end{equation}

\begin{equation}
f_{xc}(q,0) \rightarrow k^{-2}_\text{F} H\left(\frac{q}{2k_\text{F}}\right) \hspace{1in} (k_\text{F} \rightarrow 0)
\label{eq5}
\end{equation}
\noindent Table I shows the density dependences of some of the ingredients of our kernel to be introduced later. The macroscopic or slowly-varying-density limit is achieved when $\frac{q}{2k_\text{F}} \rightarrow 0$\\

Thus, in the high-density limit for fixed finite $\frac{q}{2k_\text{F}}$, $\epsilon$(q,0) $\rightarrow$ 1 + O($k^{-1}_\text{F}$) and $\chi$(q,0) $\rightarrow$ $\chi_\text{KS}$(q,0). In the adiabatic connection fluctuation dissipation expression \cite{5,6,7} for the exchange-correlation energy, the exchange energy per electron $\epsilon_x$ $\sim$ k$_\text{F}$ arises from $\chi_\text{KS}$ (as a function of wavevector and imaginary frequency), and the correlation energy per electron $\epsilon_c$ from $\chi - \chi_\text{KS}$ (as a function of coupling constant, wavevector, and imaginary frequency). The correlation energy is much smaller than the exchange energy at high densities, but tends to about 0.9$\epsilon_x$ at very low densities. In this paper, for the ingredients of f$_{xc}$, we will employ the parametrization of Ref. \cite{9} for the r$_s$ -dependence of $\epsilon_c$ at zero spin polarization.
\begin{table}
	\caption{Density (n) dependences of key ingredients (to be defined later) of the exchange-correlation kernel f$_{xc}$(q, $\omega$) for a uniform electron gas with density parameter r$_s$ $=$ $(\frac{3}{4\pi n})^{1/3}$, and of related quantities: the Fermi wavevector k$_\text{F}$ $=$ 1.9192/r$_s$, the bulk plasma frequency $\omega_p$ $=$ (4$\pi$n)$^{1/2}$, and $\Delta f_0/{f_0}$ from Ref. \cite{24}(atomic units).}
	\begin{tabular}{lrrrrrrrr}
	
	r$_s$ & k$_F$ & $\omega_p$ & k$^{-1/2}$/k$_\text{F}$ & b$^{1/2}\omega_p$ & k$^2_\text{F} f_0$ & k$^2_\text{F} f_\infty$ & k$^2_\text{F} f_{xc}(\infty,\omega)$ & $\Delta f_0/f_0$ \\
	\hline
	0 & $\infty$ & $\infty$ & 1.15 & $\infty$ & -3.14 & -1.89 & 0 & - \\
	1 & 1.92 & 1.73 & 1.67 & 0.51 & -3.25 & -1.10 & -0.39 & -0.16 \\
	2 & 0.96 & 0.61 & 1.76 & 0.49 & -3.36 & -0.92 & -0.51 & -0.12 \\
	3 & 0.64 & 0.33 & 1.80 & 0.48 & -3.45 & -0.85 & -0.57 & -0.10 \\
	4 & 0.48 & 0.22 & 1.82 & 0.47 & -3.53 & -0.83 & -0.61 & -0.08 \\
	5 & 0.38 & 0.15 & 1.83 & 0.46 & -3.60 & -0.83 & -0.63 & -0.08 \\
	$\infty$ & 0 & 0 & 2.06 & 0 & -6.07 & -3.65 & 0 & - \\
	\hline
	\end{tabular}
	
\end{table}

\section{M\MakeLowercase{odified \MakeUppercase{CP}07 static kernel}}
We begin with the static limit of the original CP07 kernel of Eq. (12) of Ref. [14] (in atomic units):
\begin{equation}
f^{CP07}_{xc}(q,0) = \left(\frac{4\pi}{q^2}\right)B[e^{-kq^2} - 1] - \left(\frac{4\pi}{k^2_\text{F}}\right)\frac{C}{[1 + \frac{1}{q^2}]},
\label{eq6}
\end{equation}

\begin{equation}
k = \frac{A}{4\pi B}.
\end{equation}
\noindent Here A, B, and C are positive functions of density n defined in Ref. \cite{14}. These functions of density typically require derivatives of  $\epsilon_{xc}(r_s)$ or $\epsilon_{c}(r_s)$, for which we employ the parametrizations from the appendix of Ref. \cite{9} (numerically almost identical to those of Refs. \cite{10} and \cite{11}) instead of the less-accurate but simpler ones of Ref. \cite{14}. CP07 is a constraint-based kernel that aims to reproduce the known small q and large q behaviors of the exact kernel:
\begin{equation}
f_{xc}(q,0) \rightarrow -A \hspace{2.2in} (q \rightarrow 0)
\label{eq8}
\end{equation}
\begin{equation}
f_{xc}(q,0) \rightarrow -\left(\frac{4\pi}{k^2_\text{F}}\right)C - \left(\frac{4\pi}{q^2}\right)B \hspace{1in} (q \rightarrow \infty).
\label{eq9}
\end{equation}
\noindent Eq. (8) is the well-known compressibility sum rule; approximating f$_{xc}$ by $-$A is the adiabatic local density approximation. Eq. (9) is from Refs. \cite{19,20}. Note that C arises from correlation alone, and vanishes in the high- and low-density limits, as shown in Table I.\\

First of all, 1/q$^2$ in Eq. (6) needs to be replaced by 1/(kq$^2$)$^2$. This substitution is needed to recover Eq. (9) and the density scalings discussed in section 2 of this article. Note that, by Table 1, kq$^2$ scales like ($\frac{q}{2k_\text{F}}$)$^2$ in the high- and low-density limits.\\

The second change we make is to replace Eq. (8) by the more-detailed
\begin{equation}
f_{xc}(q,0) \rightarrow -A + Dq^2 \hspace{1in} (q \rightarrow 0),
\label{eq10}
\end{equation}
\begin{equation}
D = \frac{2C_{xc}(r_s)}{n^{4/3}} \hspace{1.9in}
\end{equation}
\noindent from Eqs. (25), (32), and (37) of Ref. \cite{18}, but with improved input. The q $\rightarrow$ 0 limit is the limit of slowly-varying-in-space induced density, in which the second-order gradient expansion becomes exact. Thus, in Eq. (11), C$_{xc}(r_s)$ is the coefficient of the second-order gradient expansion for the exchange-correlation energy:
\begin{equation}
C_{xc}(r_s) = C_x + C_c(r_s=0)\frac{1 + 3.138r_s + 0.3r^2_s}{1 + 3.0r_s + 0.5334r^2_s},
\label{eq12}
\end{equation}
\noindent with C$_x =$ $-$ 0.00238 \cite{29} and C$_c (r_s = 0) =$ 0.00423 \cite{30}. Untypically, C$_{xc}$ does not reduce to C$_x$ as r$_s \rightarrow$ 0. We have used the r$_s$ - dependence of Eq. (36) of Ref. \cite{31}, in which C$_{xc}$ decreases very slowly to zero as r$_s$ increases, taking the values 0.00185, 0.00122, and 0.00015 at r$_s =$ 0, 4, and 70, respectively. This means that, at very low densities with r$_s \geq$ 70, D will be close to 0 and the LDA kernel will be nearly correct through order $(\frac{q}{2k_\text{F}})^2$.\\

A better match to the QMC kernel \cite{19} for r$_s$ in the metallic range and for $\frac{q}{2k_\text{F}}$ $< \hspace{0.2cm} \sim$ 1 could be achieved by setting D $=$ 0 in Eq. (10). Within its error bars, the QMC kernel can also be matched \cite{32} by including higher-order terms in the gradient expansion of the exchange-correlation energy, although the fourth-order terms are not known for the correlation energy. In the interests of simplicity and generality, we have not included a q$^4$ term in Eq. (10). The goal of constraint satisfaction is not to match every detail, but to make a correct global map.\\

The result of these changes is the modified CP07 (MCP07) static kernel:
\begin{equation}
f^{MCP07}_{xc}(q,0) = \left(\frac{4\pi}{q^2}\right) B [e^{-kq^2}(1 + Eq^4) - 1] - \left(\frac{4\pi}{k^2_\text{F}}\right)\frac{C}{[1 + \frac{1}{(kq^2)^2}]},
\label{eq13}
\end{equation}

\begin{equation}
E = \frac{D}{4\pi B} - \frac{k^2}{2}.
\label{eq14}
\end{equation}
\noindent Its exchange-only and exchange-correlation incarnations for r$_s =$ 4 are plotted in Fig. 1. We see that, away from q $=$ 0, correlation can be more important than exchange.\\

We use the name MCP07 only for the static limit of our kernel, since our full kernel will also modify the Gross-Kohn expression.

\section{S\MakeLowercase{tatic charge-density wave in jellium}}
Overhauser \cite{17} proposed that periodic metals could be unstable against the
formation of a static charge-density wave (CDW). Quantum Monte Carlo calculations found a CDW or incipient body-centered cubic (bcc) Wigner crystallization in spin-polarized jellium at a low critical density corresponding to r$_s =$ 70 \cite{33}, or at r$_s = 85 \pm 20$ in spin-unpolarized jellium \cite{8}. The 1980
calculation of Ceperley and Alder \cite{8} also found that ground-state jellium remains
spin-unpolarized for r$_s \leq 75 \pm 5$. In the same year, Perdew and Datta \cite{18}, using a static kernel designed to satisfy Eqs. (10) and (11), also found a CDW near this critical r$_s$ .\\

Figure 2 of the present article, which is similar to Fig. 4 of Ref. \cite{18}, was
found by fixing a value for $\frac{q}{2k_\text{F}}$ and then searching for the largest value of k$_\text{F}$ that makes $\epsilon(q) =$ 0 (hence a non-zero density response at wavevector q even in the absence of any perturbing potential). This happens around r$_s =$ 30 in the adiabatic local density approximation, and around r$_s =$ 69 in MCP07. We could not find a charge-density wave at any density from the original CP07. All the low-density instabilities of jellium are difficult to pinpoint, because the energies of the different phases as functions of r$_s$ are nearly the same at low densities.\\

The charge-density wave first appears with $\frac{q}{2k_\text{F}} \approx$ 1.14, making q close to the first reciprocal lattice vector of a bcc Wigner crystal with one electron per primitive cell. For much smaller $\frac{q}{2k_\text{F}}$, the CDW is strongly suppressed by the Coulomb term 4$\pi/q^2$ in Eq. (2). Later in this article, we will show that the CDW is associated with a softening of the plasmon mode.\\

Our Fig. 1 shows that the static MCP07 kernel f$_{xc}$(q,0) is always more
negative than its exchange-only version f$_x$(q,0). This result confirms Overhauser's 1968 prediction that correlation enhances the charge-density wave

\section{F\MakeLowercase{requency-dependent local kernel of \MakeUppercase{G}ross and \MakeUppercase{K}ohn}}
A constraint-based model for f$_{xc}$(q=0,$\omega$) was proposed in 1985 by Gross and Kohn \cite{2}, and later corrected by Iwamoto and Gross \cite{15}. It starts from a
constrained interpolation for the imaginary part, evaluated at a real frequency,
between known real zero- (f$_0$) and infinite- (f$_\infty$) frequency limits at q $=$ 0:
\begin{equation}
Im f_{xc}(0,\omega) = -cb^{3/4}g(b^{1/2}\omega),
\label{eq15}
\end{equation}
\begin{equation}
g(x) = \frac{x}{(1 + x^2)^{5/4}},
\label{eq16}
\end{equation}
\begin{equation}
b = \{\left(\frac{\gamma}{c}\right)[f_\infty - f_0]\}^{4/3},
\label{eq17}
\end{equation}
\begin{equation}
c = 23\frac{\pi}{15},
\label{eq18}
\end{equation}
\begin{equation}
\gamma = \frac{\left[\Gamma\left(\frac{1}{4}\right)\right]^2}{(32\pi)^{1/2}} = 1.311.
\label{eq19}
\end{equation}

\noindent Figure 3 shows this function of real $\omega$ for r$_s =$ 4, and also its exchange-only contribution. Again the importance of correlation is manifest. Table I shows that the dimensionless quantity b$^{1/2}\omega_p$ (where $\omega_p = (4\pi n)^{1/2}$ is the bulk plasmon frequency) is nearly constant over the range of metallic densities, but not over all densities. Thus, in the metallic range, g(b$^{1/2} \omega$) is approximately a function of $\frac{\omega}{\omega_p}$.\\

The next step is to use the Kramers-Kronig relations \cite{2} between the imaginary and real parts of f$_{xc}$ - f$_\infty$ at real frequency to find
\begin{equation}
Re \hspace{0.1cm} f_{xc}(0,\omega) - f_\infty = \left(\frac{1}{\pi}\right) \hspace{0.1cm} P \int_{-\infty}^{\infty} d\omega' \hspace{0.1cm} \frac{Im f_{xc}(0,\omega')}{\omega' - \omega}.
\label{eq20} 
\end{equation}
\noindent As $\omega \rightarrow \infty$, $Im f_{xc}(0,\omega) \sim -c/\omega^{3/2}$ and $Re f_{xc}(0,\omega) - f_\infty \sim c/\omega^{3/2}$.\\

\noindent The principal value of the integral can be found numerically. However, the scaling relation of Eq. (15) implies the scaling relation
\begin{equation}
Re f_{xc}(0,\omega) - f_\infty = -c \hspace{0.1cm}b^{3/4} \hspace{0.2cm}h(b^{1/2}\omega),
\label{eq21}
\end{equation}
\noindent where h(0) $= \frac{1}{\gamma}$  to recover the correct non-zero $\omega \rightarrow 0$ limit. A fair fit with the correct large-$\omega$ asymptotics is provided by the simple algebraic model
\begin{equation}
h_{model}(x) = \frac{(\frac{1}{\gamma})[1 - ax^2]}{[1 + (a/\gamma)^{4/7}x^2]^{7/4}}.
\label{eq22}
\end{equation}
\noindent with a fit parameter a $=$ 0.63.\\

Figure 4 compares the real-frequency dependences of the real part of the
kernel, with and without correlation, from the Kramers-Kronig relation and from
the model. The model is less accurate at intermediate frequencies than at low or
high frequencies. While the Kramers-Kronig choice is the consistent one, we have
found that it does not make any significant difference from the model in the
applications presented here.\\

For calculation of the correlation energy, we will need f$_{xc} (0,\omega)$ for
frequencies $\omega$ in the upper-half complex plane, where this function is analytic \cite{2}. For this, we use the Cauchy integral over real $\omega'$:
\begin{equation}
f_{xc}(0,\omega) - f_\infty = \frac{1}{2\pi i} \int_{\infty}^{\infty} d\omega' \frac{[f_{xc}(0,\omega') - f_\infty]}{\omega' - \omega}.
\label{eq23}
\end{equation}
\noindent By letting $\omega$ approach the real axis from above, we can derive the Kramers-Kronig relations including Eq. (20). Figure 5 shows the kernel for frequencies on the upper imaginary axis, where the kernel is purely real.

\section{C\MakeLowercase{ombining the wavevector dependence of} MCP07 \MakeLowercase{with the frequency dependence of the} G\MakeLowercase{ross}-K\MakeLowercase{ohn kernel}}
An important constraint is Eq. (5.176) of Ref. \cite{22}, attributed there to Ref. \cite{34}. It says that the $\omega$-dependence of the kernel damps out at large q, even when the kernel itself has a non-zero large-q limit.\\

To avoid empiricism, we will use the same Gaussian damping factor that damps
out the local density and gradient expansion terms at large q in Eq. (13):
\begin{equation}
f_{xc}(q,\omega) = \left[1 + e^{-kq^2}\left\{\frac{f_{xc}(0,\omega)}{f_{xc}(0,0)} - 1\right\}\right]f^{MCP07}_{xc}(q,0).
\label{eq24}
\end{equation}
\noindent When q $=$ 0, Eq. (24) properly recovers f$_{xc}$(0,$\omega$). When $\omega = 0$, Eq. (24) properly recovers f$^{MCP07}_{xc}(q,0)$. And when q$\rightarrow \infty$, Eq. (24) correctly reduces to f$^{MCP07}_{xc}(q,0)$.\\

Figure 6 shows the q-dependence of the imaginary part of Eq. (24) for r$_s =$
4 for various real frequencies that are integer multiples of the bulk plasmon
frequency. Figure 7 shows the same for the real part (using the model of Eq. (22)). Note that the frequency dependence is already strongly damped at $\frac{q}{2k_\text{F}} = 1$.\\

A viscosity correction to the compressibility value for f$_0 = f_{xc}$(0,0) was
found by Conti and Vignale \cite{35}. It is of order 10\% at metallic densities, as shown by the values of $\Delta f_0/f_0$ in Table I (based on $\Delta f_0$ values from Ref. \cite{24}), and is not
included in our Eq. (24).

\section{P\MakeLowercase{lasmon in jellium}}

The plasmon is a collective long-wavelength oscillation of the electron
density, at a frequency $\omega_p(q)$ that tends as q $\rightarrow$ 0 to the classical limit or bulk plasmon frequency $\omega_p = (4\pi n)^{1/2}$. At q less than a critical wavevector q$_c$, the real part of the complex plasmon energy $\omega_p(q)$ lies above the highest energy of the continuum of single electron-hole excitations of wavevector q, which in a non-
interacting picture has a highest energy of $\frac{(k_\text{F} + q)^2}{2} - \frac{k^2_\text{F}}{2}$. Thus
\begin{equation}
\frac{Re \hspace{0.2cm} \omega_p(q_c)}{k^2_\text{F}} = \left(\frac{1}{2}\right) \left(\frac{q_c}{k_\text{F}}\right)^2 + \frac{q_c}{k_\text{F}}.
\label{eq25}
\end{equation}
\noindent In the range q $< q_c$ (the only range we will consider here), the plasmon excitation cannot decay to a single electron-hole pair excitation, so its lifetime is infinite for any real (hence frequency-independent kernel). But a frequency-dependent kernel should yield a plasmon frequency in the lower-half complex frequency plane, where Im $\omega_p(q)$ is minus the inverse of a lifetime arising from decay of the plasmon into multiple electron-hole pairs.\\

We find $\omega_p(q)$ by fixing a real wavevector q and searching over complex
frequencies $\omega$ for the one that zeroes out $\epsilon(q,\omega)$ of Eq. (2). In practice, we stop when $\mid$$\epsilon$$\mid$ is of order 10$^{-3}$. Since our Cauchy integral of Eq. (23) is only for $\omega$ in the
upper-half complex plane, we find f$_{xc}(0,\omega)$ by analytic continuation or Taylor
expansion from a near frequency on the real axis. The zero-th order term of this
expansion almost suffices, as we confirm by adding the first-order term, using the
Cauchy-Riemann conditions on an analytic function to convert known derivatives
of the real and imaginary parts of f$_{xc}$ with respect to Re $\omega$ to derivatives with
respect to Im $\omega$.\\

Figure 8 shows the resulting plasmon dispersion or Re $\omega_p(q)$ for r$_s =$ 4, which would be almost the same if the frequency dependence of the kernel were neglected, and not qualitatively different if the kernel were set to 0. Fig. 9 shows the resulting Im $\omega_p(q)$, or minus the inverse plasmon lifetime, which would equal
zero without the frequency dependence. The calculated inverse lifetime grows like
q$^2$ at small q, as expected \cite{22}, but starts to decrease again as q approaches k$_\text{F}$, where the Gross-Kohn frequency dependence is increasingly damped out via our
Eq. (24). The minimum predicted plasmon lifetime is of the order of femtoseconds.\\

Figure 10 shows Re $\omega_p(q)$ for r$_s =$ 69, where the static charge density wave
was found to appear in section 4. Unlike the dispersion in Fig. 8, the dispersion here
is downward, and $\omega_p(q)$ appears to be heading toward zero at q/k$_\text{F} \approx$ 2. Thus the static charge density wave can be understood to arise from a soft plasmon mode.

\section{C\MakeLowercase{orrelation energy per electron in jellium}}
The correlation energy of the uniform electron gas has a long history, going back to
the random phase approximation (RPA) (f$_{xc} = 0$) of the 1950's \cite{5}. The formula
we use here is Eq. (27) of Ref. \cite{27}. In this equation, an integral over real
frequencies from 0 to $\infty$ has been transformed by contour integration to an integral
over imaginary frequencies in the upper half plane. This is done to avoid the
plasmon pole near the real axis, and results in the smooth frequency integrand
shown in Figs. 3 and 4 of Ref. \cite{27}. The Kohn-Sham non-interacting and real
interacting systems are connected adiabatically through the coupling constant λ
between 0 and 1 in the Coulomb interaction $\frac{4\pi \lambda}{q^2}$. The exchange-correlation kernel
must also be scaled, as in Eq. (18) of Ref. \cite{27}:
\begin{equation}
f^\lambda_{xc}(n,q,\omega) = \lambda^{-1}f_{xc}(\frac{n}{\lambda^3},\frac{q}{\lambda},\frac{\omega}{\lambda^2}).
\end{equation}

\noindent Figure 11 shows our results for the correlation energy per electron as a function of r$_s$ in the metallic range, in comparison with the highly-accurate parametrization and extension \cite{13} of the results of Ref. \cite{8} by Perdew and Wang 1992 \cite{11} (indistinguishable on the scale of the figure from the parametrization of
Ref. \cite{9}). As is well known, RPA (f$_{xc} \rightarrow$ 0) makes the correlation energy per electron too low by about 0.4 eV/electron, and the adiabatic local density
approximation (f$_{xc} \rightarrow$ f$_{xc}$(0,0)) makes it too high by about the same absolute error. A good kernel f$_{xc}$(q,$\omega$) should produce an accurate result, and our static MCP07 kernel does so to a remarkable extent. Adding the frequency dependence of Eq. (24) degrades the accuracy, but almost negligibly. Replacing the Gaussian in
Eq. (24) by 1 (thus using the undamped Gross-Kohn frequency dependence) would degrade the accuracy significantly, correcting only about 2/3 of the RPA error.

\section{C\MakeLowercase{onclusions}}
The CP07 exchange-correlation kernel for zero frequency and the Gross-Kohn
kernel for zero wavevector were constructed for the uniform electron gas via the satisfaction of exact constraints. By imposing further exact constraints, we have made an improved MCP07 static kernel and combined it with the Gross-Kohn dynamic kernel. Key added constraints include the second-order gradient expansion for the exchange-correlation energy, and the damping out of the frequency dependence with increasing wavevector. That damping out is already substantial at q $\approx$ k$_\text{F}$.\\

Without any fitting, we have achieved high accuracy for all studied
properties. In particular, the critical density (r$_s \approx$ 69) and critical wavevector of the static charge-density wave that appears at low density are accurate. We have shown that this ground-state instability of the uniform phase is associated with a soft plasmon. We have also found that correlation enhances the instability, as Overhauser predicted in 1968. That is evident at r$_s =$ 4 from Fig. 2, and is expected to remain true at larger r$_s$ where correlation becomes relatively more important.\\

We have studied the plasmon at the density of metallic sodium (r$_s =$ 4),
where our frequency dependence produces a plasmon lifetime that first decreases from infinity to a few femtoseconds and then increases, as the wavevector increases from 0 toward the Fermi wavevector k$_\text{F}$. The increase is neither confirmed nor disconfirmed by other calculations, to our knowledge.\\

We have also calculated remarkably accurate correlation energies per
electron for metallic r$_s$ from 1 to 6. The improvement over RPA arises from the wavevector dependence of the MCP07 kernel. The frequency dependence of our kernel has almost no effect on the ground-state energy, a conclusion that might extend to real systems. Unlike jellium, real systems are known to require a long-range kernel, with f$_{xc} \sim \frac{1}{q^2}$ for q $\rightarrow$ 0, which might quantitatively correct the qualitatively-right RPA description of long-range van der Waals interaction.\\

It should be noted that \cite{36}, for metallic densities, the range of wavevectors relevant to the plasmon outside the electron-hole continuum, 0 $\leq$ q $< \hspace{0.2cm} \sim$ k$_\text{F}$, is
different from the range relevant to the correlation energy, 0 $< \hspace{0.2cm} \sim$ q $< \hspace{0.2cm} \sim$ 3k$_\text{F}$ . At
much lower densities, the latter range is also relevant to the plasmon and charge-density wave. In the range 0 $\leq$ q $< \hspace{0.2cm} \sim$ k$_\text{F}$, the effective interaction $\frac{4\pi}{q^2}$+f$_{xc}$ in Eq. (2) is dominated by $\frac{4\pi}{q^2}$ \cite{36}, so
modest deviations of f$_{xc}$ from its RPA value 0, or better from the adiabatic local density approximation f$_{xc}$(0,0), have almost negligible effect in that range, apart from emergent phenomena like the plasmon lifetime. Thus our applications test the kernel not only over a wide range of densities but also over a fairly wide range of $\frac{q}{2k_\text{F}}$. In the future, we hope to find more demanding tests for the frequency dependence.\\

In our uniform-gas exchange-correlation kernel, the full Gross-Kohn frequency dependence is unveiled only in the long-wavelength (q $\rightarrow$ 0) limit, in which the kernel itself is overwhelmed by the Coulomb interaction $\frac{4\pi}{q^2}$.  Inhomogeneous ground states \cite{22} have ultra-nonlocal kernels with a $\frac{1}{q^2}$ variation in this limit that strongly affects optical absorption, and might also have a frequency dependence.

\noindent \textbf{Acknowledgments:} AR and NKN acknowledge support from the U.S. National Science Foundation under Grant No. DMR-1553022. JPP acknowledges support from the U.S. National Science Foundation under Grant No. DMR-1939528 (CMMT-Division of Materials Theory, with a contribution from CTMC-Division of Chemistry).

\begin{figure}[h!]
	\includegraphics[scale=0.45]{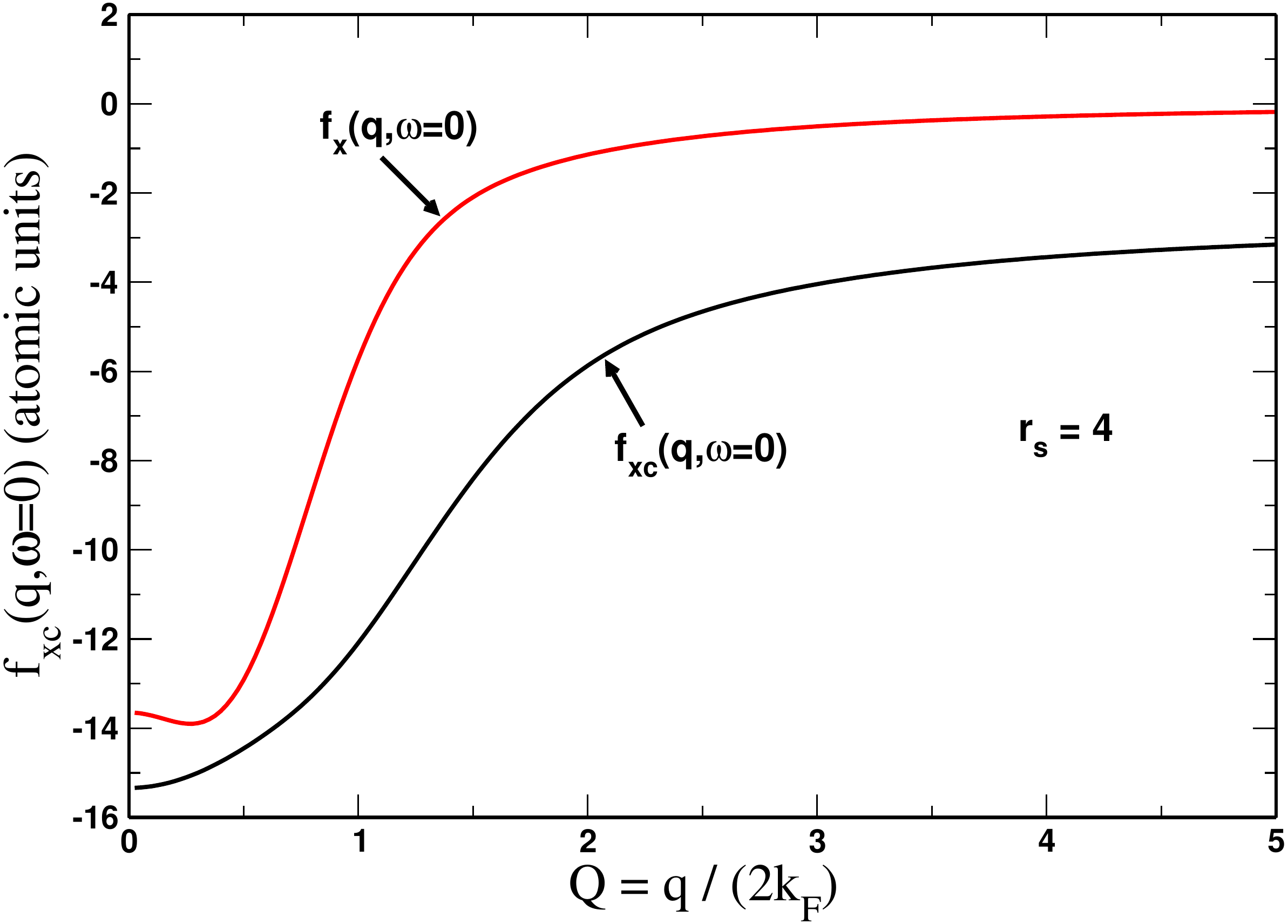}
	\caption{Modified CP07 (MCP07) static kernels for jellium with density parameter r$_s=$ 4 at the exchange-only and exchange-correlation levels, versus reduced wavevector.}
\end{figure}

\begin{figure}[h!]
	\includegraphics[scale=0.45]{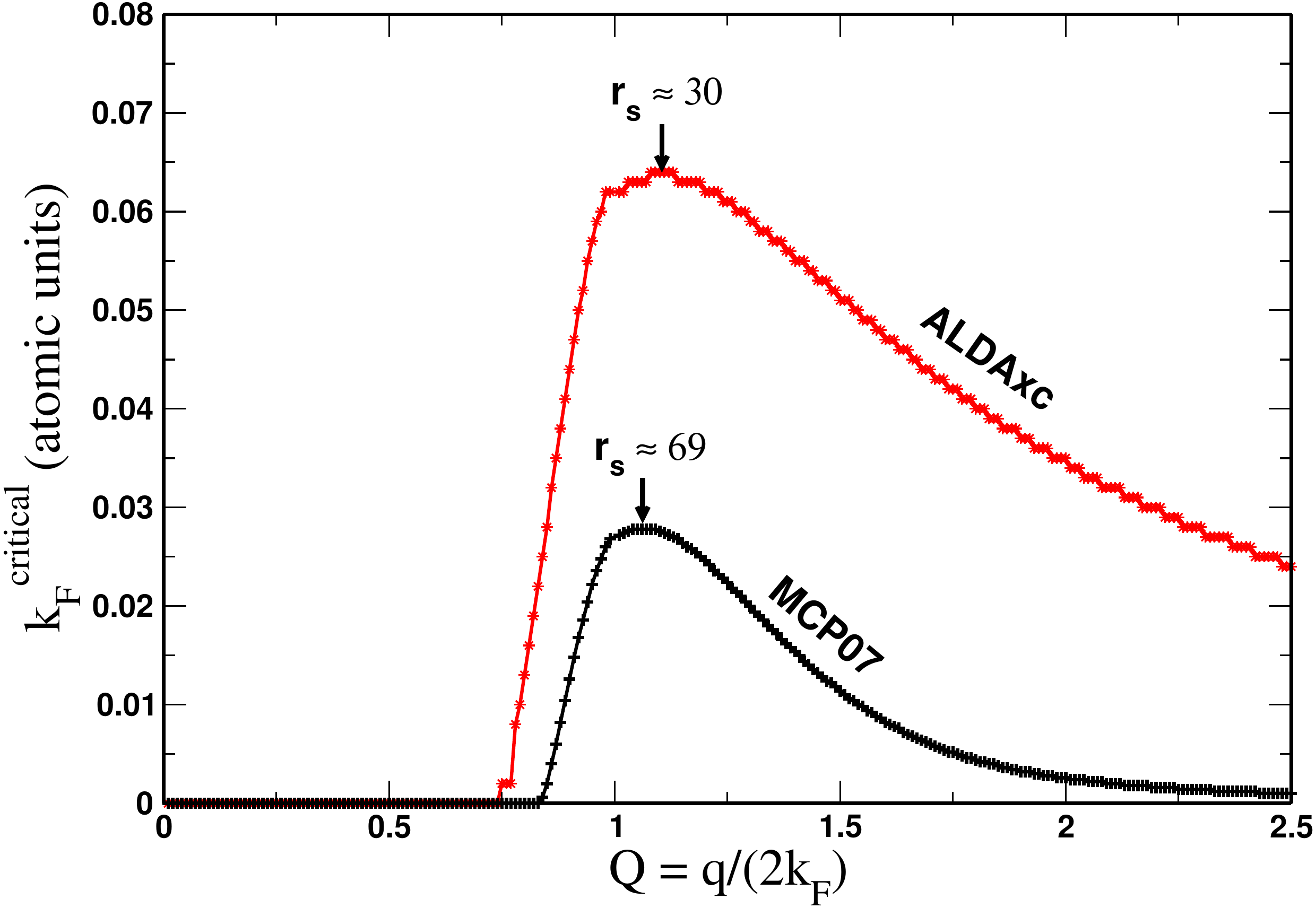}
	\caption{Critical Fermi wavevector for the appearance of a static charge-density
		wave in a low-density jellium, from the adiabatic local density approximation (f$_{xc} =$ f$_{xc}$(0,0)) and
		MCP07 static (f$_{xc} =$ f$_{xc}$(q,0)) exchange-correlation kernels, versus reduced wavevector.}
\end{figure}

\begin{figure}[h!]
	\includegraphics[scale=0.47]{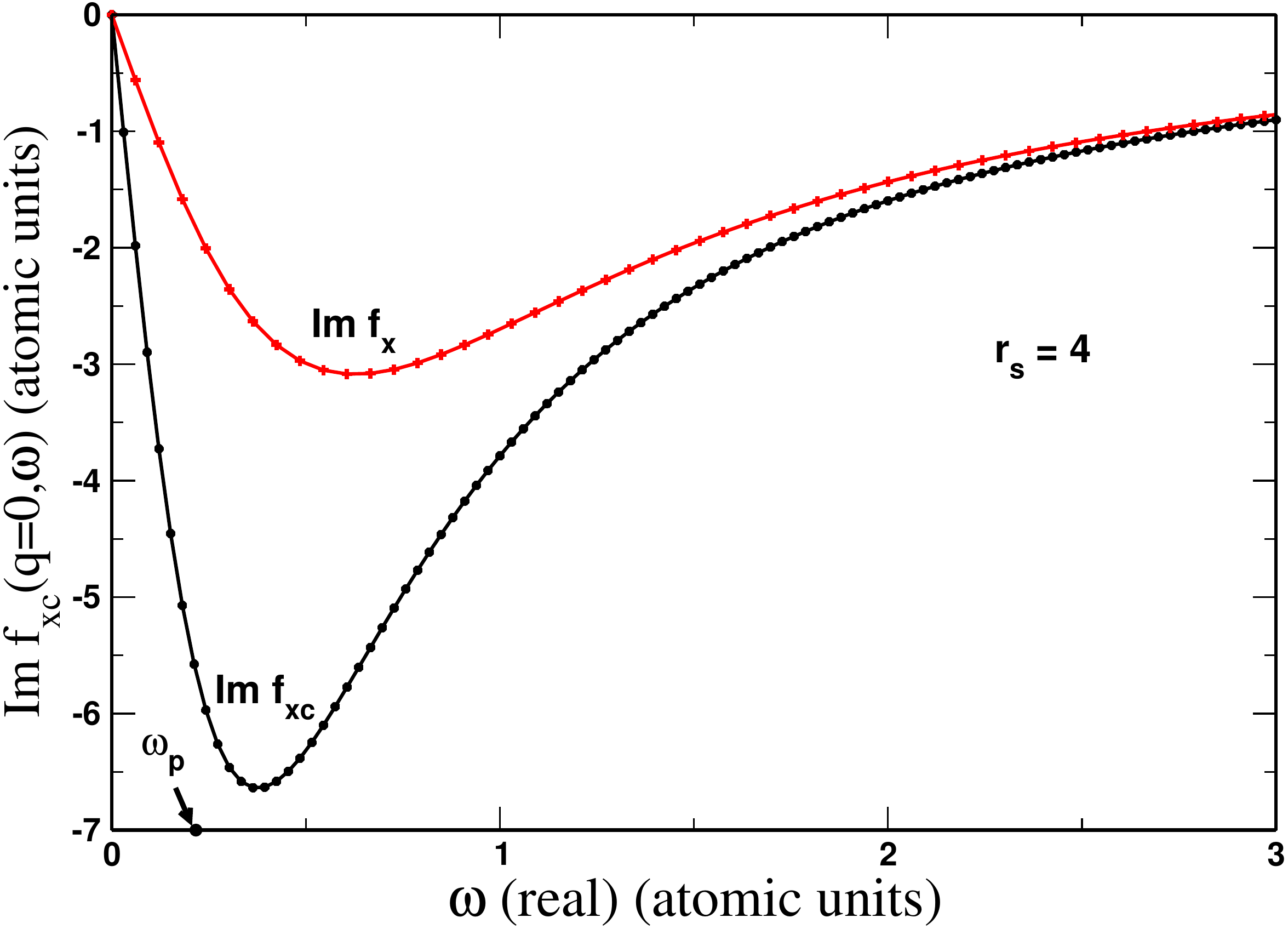}
	\caption{Imaginary part of the Gross-Kohn q $=$ 0 dynamic kernel for jellium with
		density parameter r$_s =$ 4 , at the exchange-only and exchange-correlation levels,
		versus real frequency.}
\end{figure}

\begin{figure}[h!]
	\includegraphics[scale=0.47]{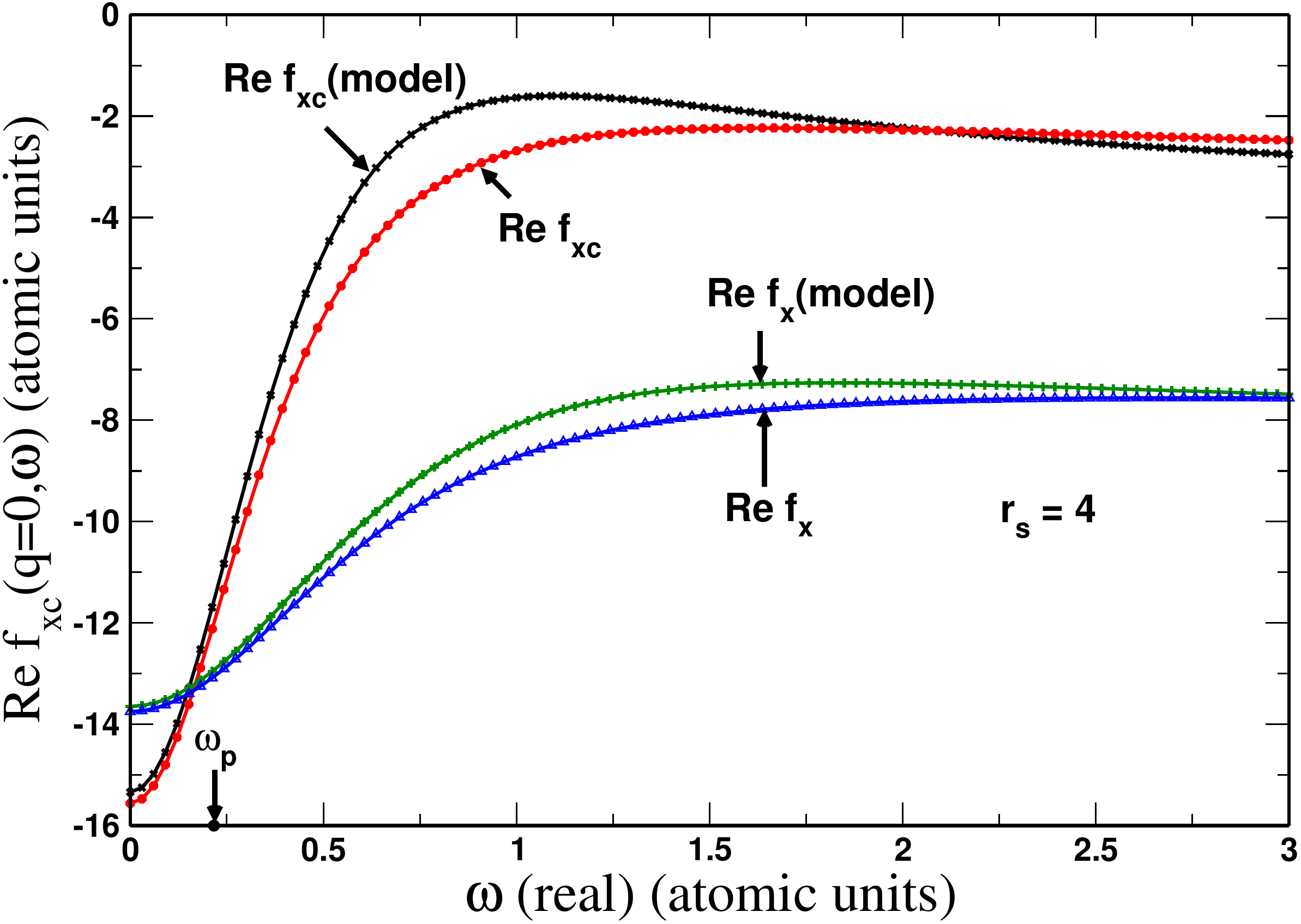}
	\caption{Real part of the Gross-Kohn q $=$ 0 dynamic kernel for jellium with density
		parameter r$_s =$ 4 , at the exchange-only and exchange-correlation levels, versus
		real frequency. (From the Kramers-Kronig relation of Eq. (20) and from the model
		of Eq. (22).)}
\end{figure}

\begin{figure}[h!]
	\includegraphics[scale=0.45]{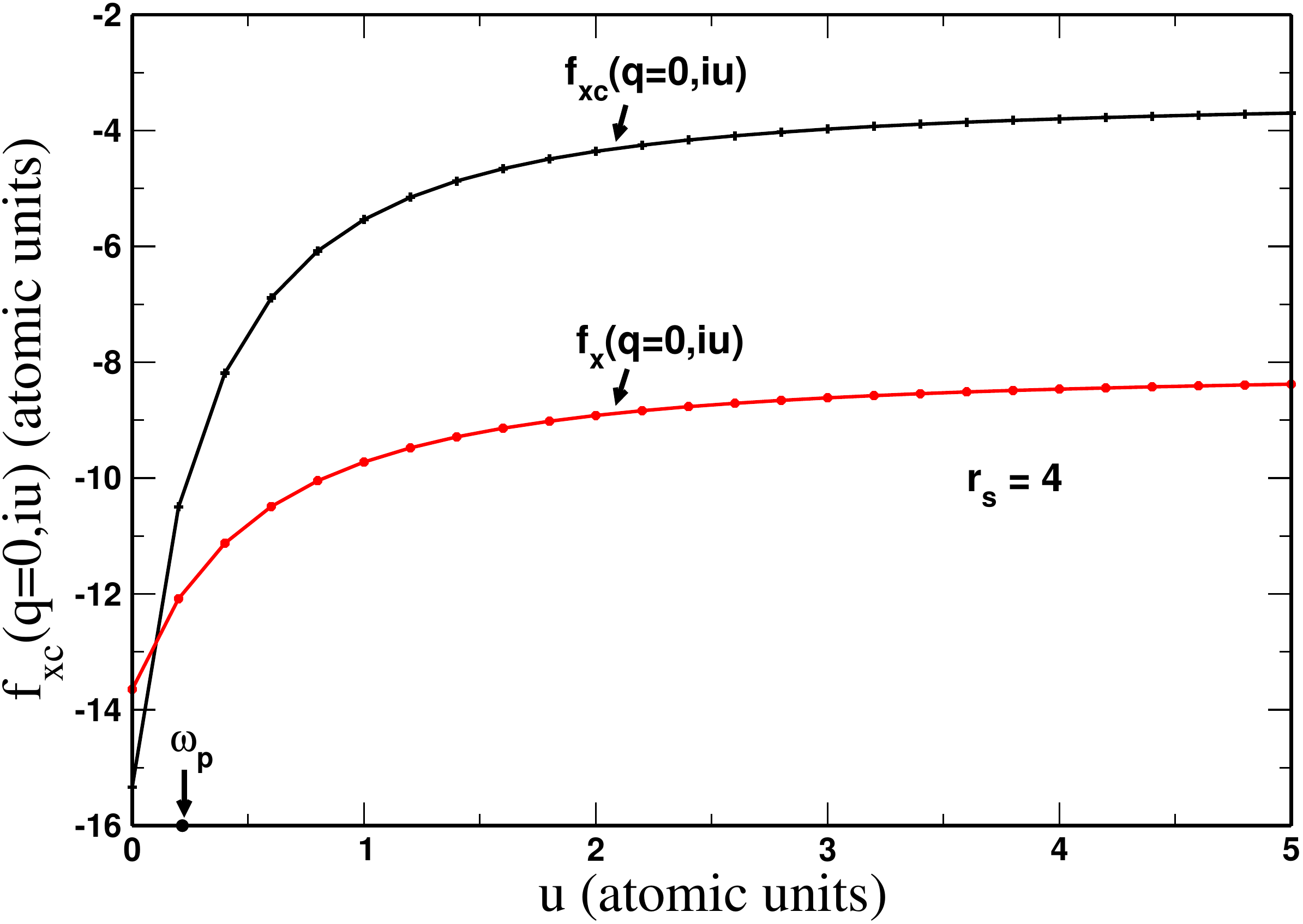}
	\caption{The purely-real Gross-Kohn q $=$ 0 dynamic kernel for jellium with density
		parameter r$_s =$ 4 , at the exchange-only and exchange-correlation levels, versus
		imaginary frequency. (From the Cauchy integral of Eq. (23) and the model of Eq.
		(22).)}
\end{figure}

\begin{figure}[h!]
	\includegraphics[scale=0.46]{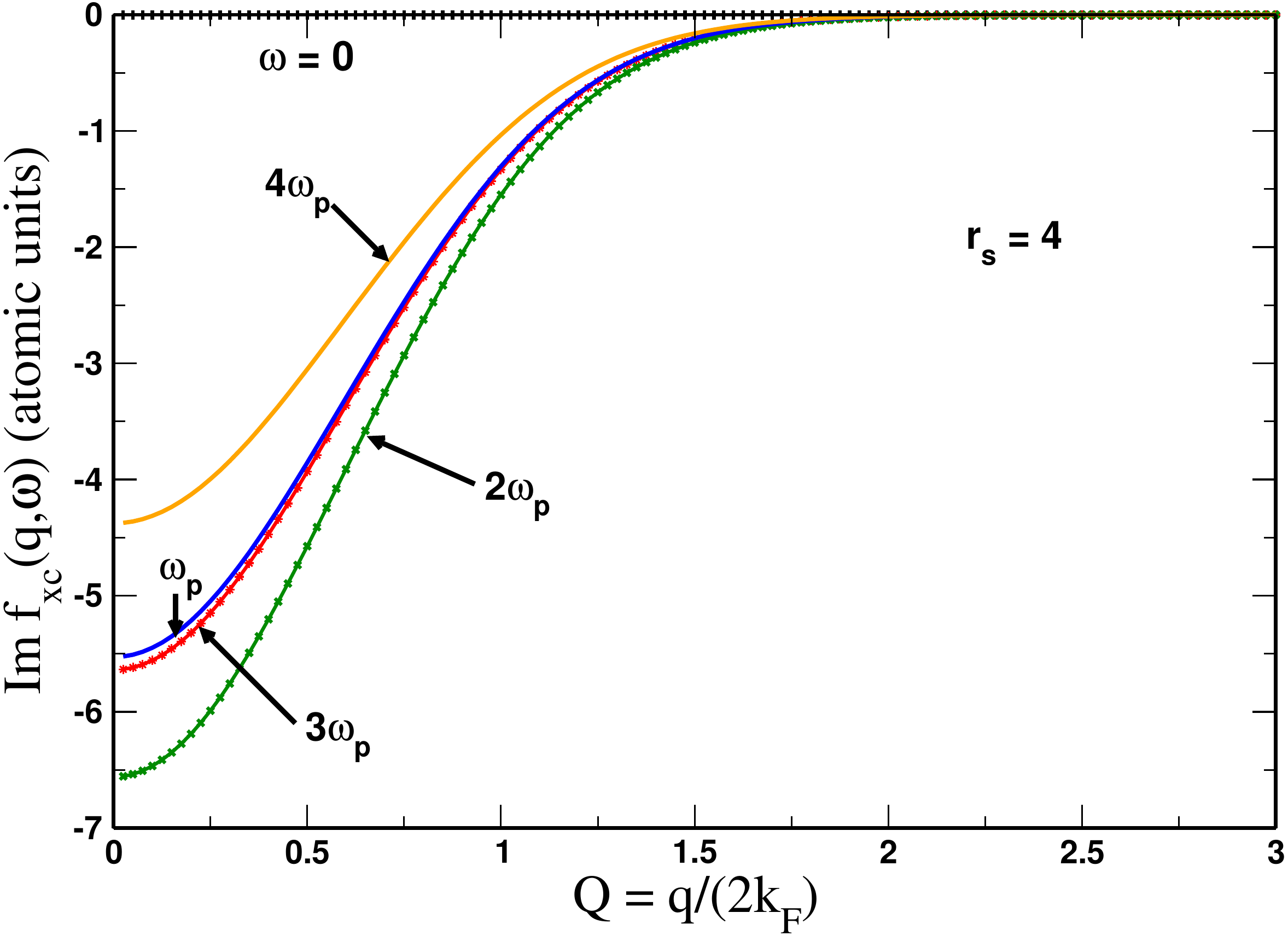}
	\caption{Imaginary part of the exchange-correlation kernel of Eq. (24) for jellium
		with density parameter r$_s =$ 4, for five different real frequencies, versus reduced wavevector.}
\end{figure}

\begin{figure}[h!]
	\includegraphics[scale=0.49]{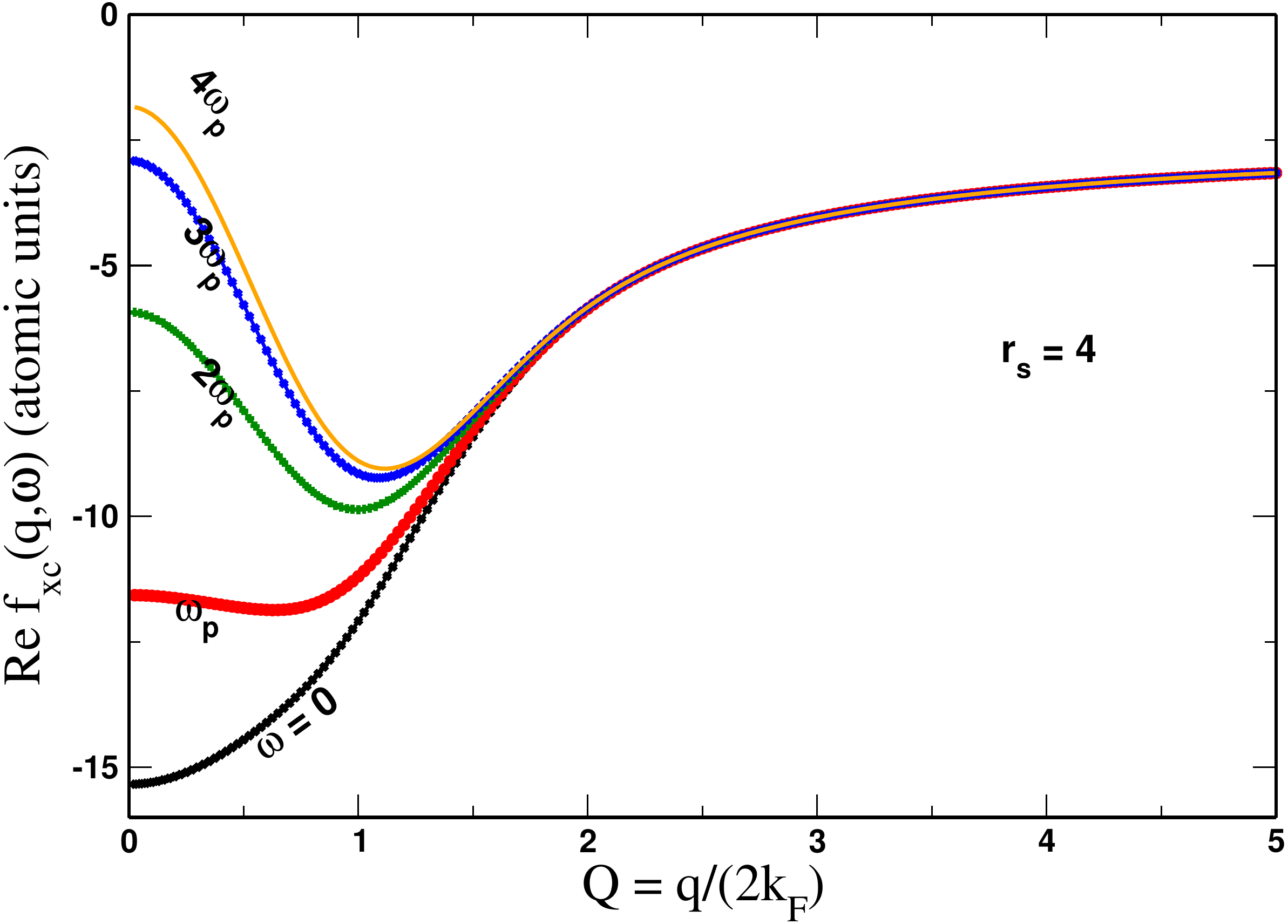}
	\caption{Real part of the exchange-correlation kernel of Eq. (24) for jellium with
		density parameter r$_s =$ 4, for five different real frequencies, versus reduced
		wavevector. (From the model of Eq. (22).)}
\end{figure}

\begin{figure}[h!]
	\includegraphics[scale=0.49]{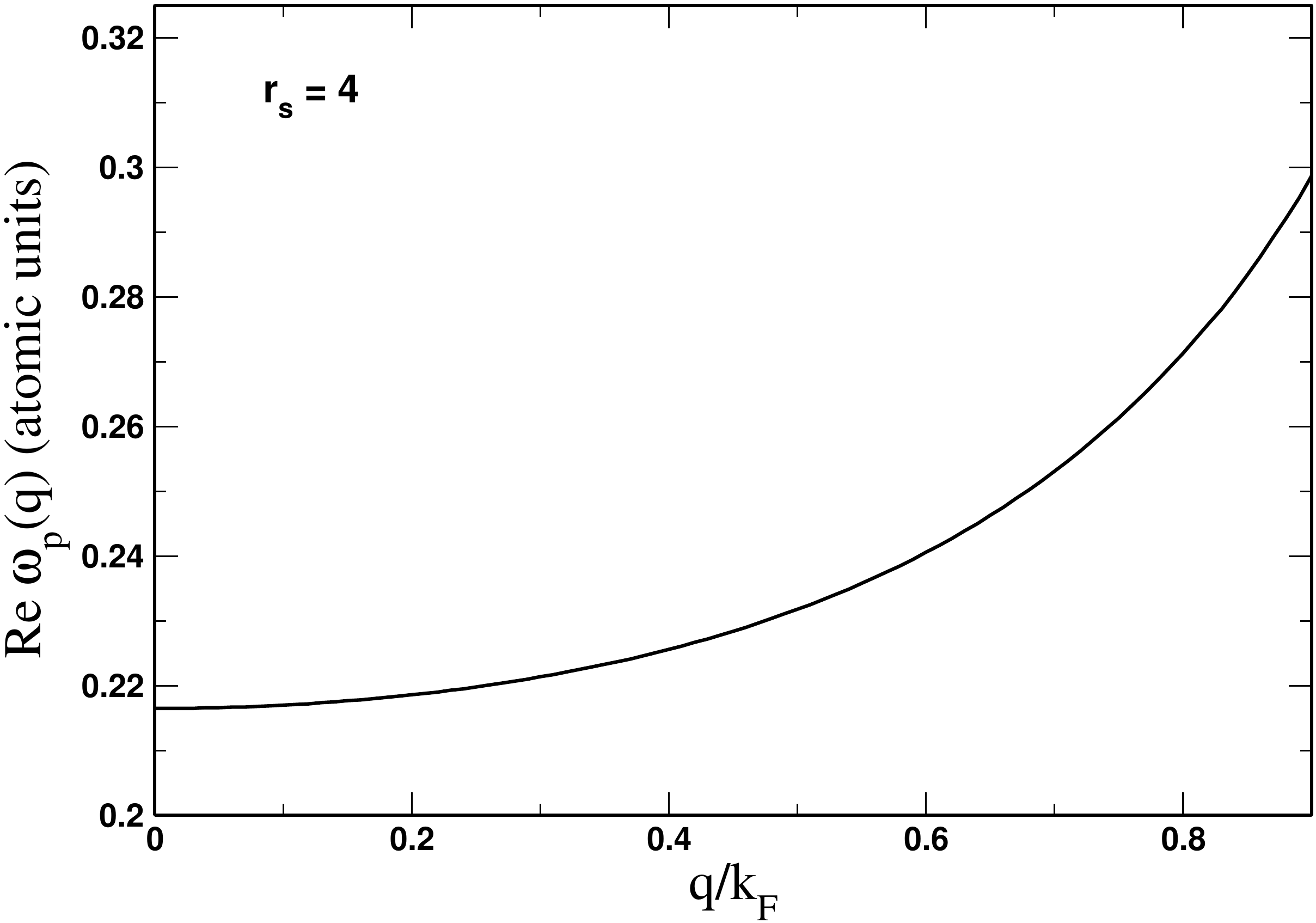}
	\caption{Plasmon dispersion for jellium with density parameter r$_s =$ 4, from the
		kernel of Eq. (24), versus reduced wavevector. The real part of the plasmon
		frequency is plotted.}
\end{figure}

\begin{figure}[h!]
	\includegraphics[scale=0.49]{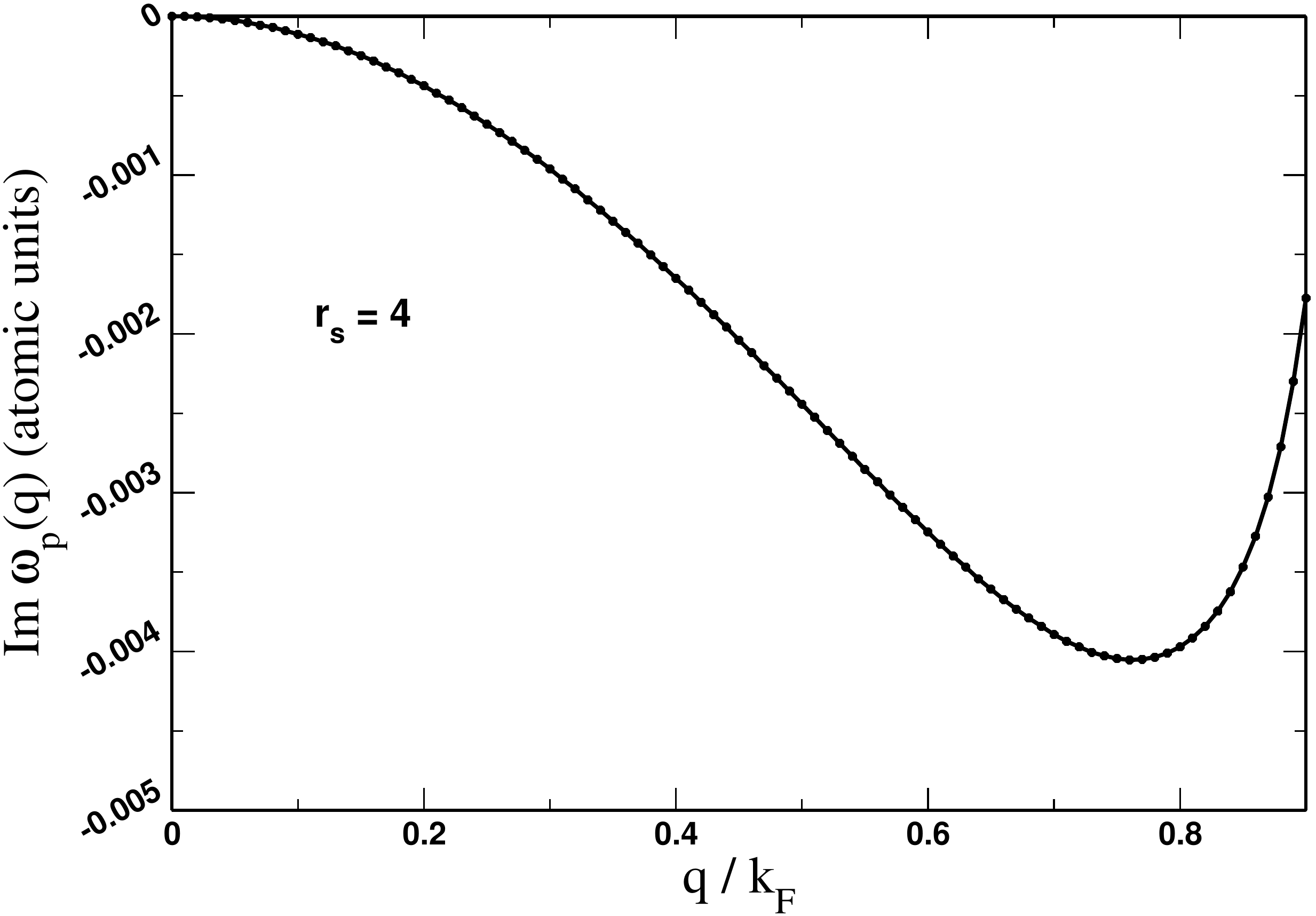}
	\caption{Plasmon damping for jellium with density parameter r$_s =$ 4 , from the
		kernel of Eq. (24), versus reduced wavevector. The imaginary part of the plasmon
		frequency is plotted.}
\end{figure}

\begin{figure}[h!]
	\includegraphics[scale=0.46]{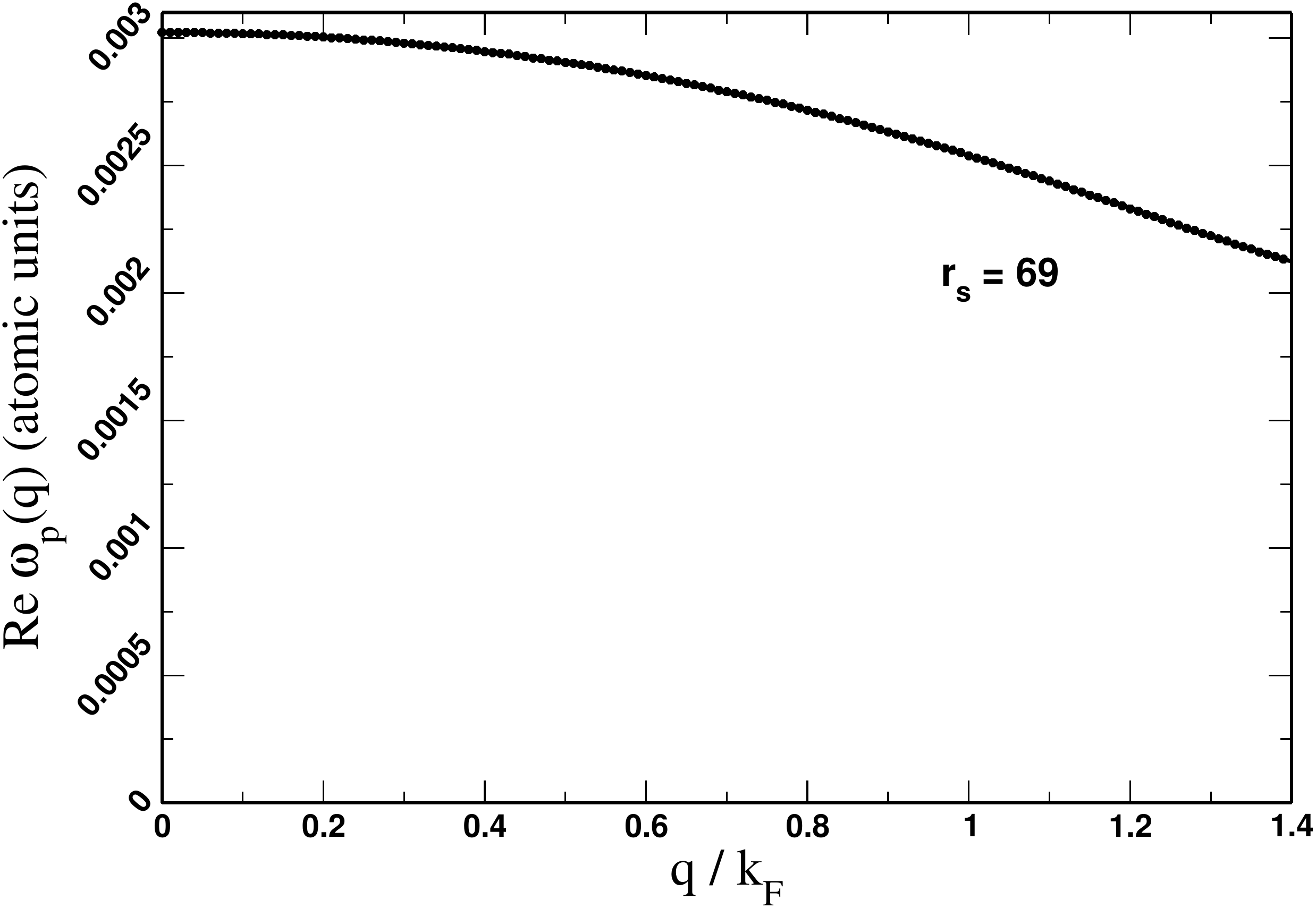}
	\caption{Plasmon dispersion for jellium with density parameter r$_s =$ 69, from the
		kernel of Eq. (24), versus reduced wavevector. The softened plasmon mode may
		lead to the static charge-density wave.}
\end{figure}

\begin{figure}[h!]
	\includegraphics[scale=0.44]{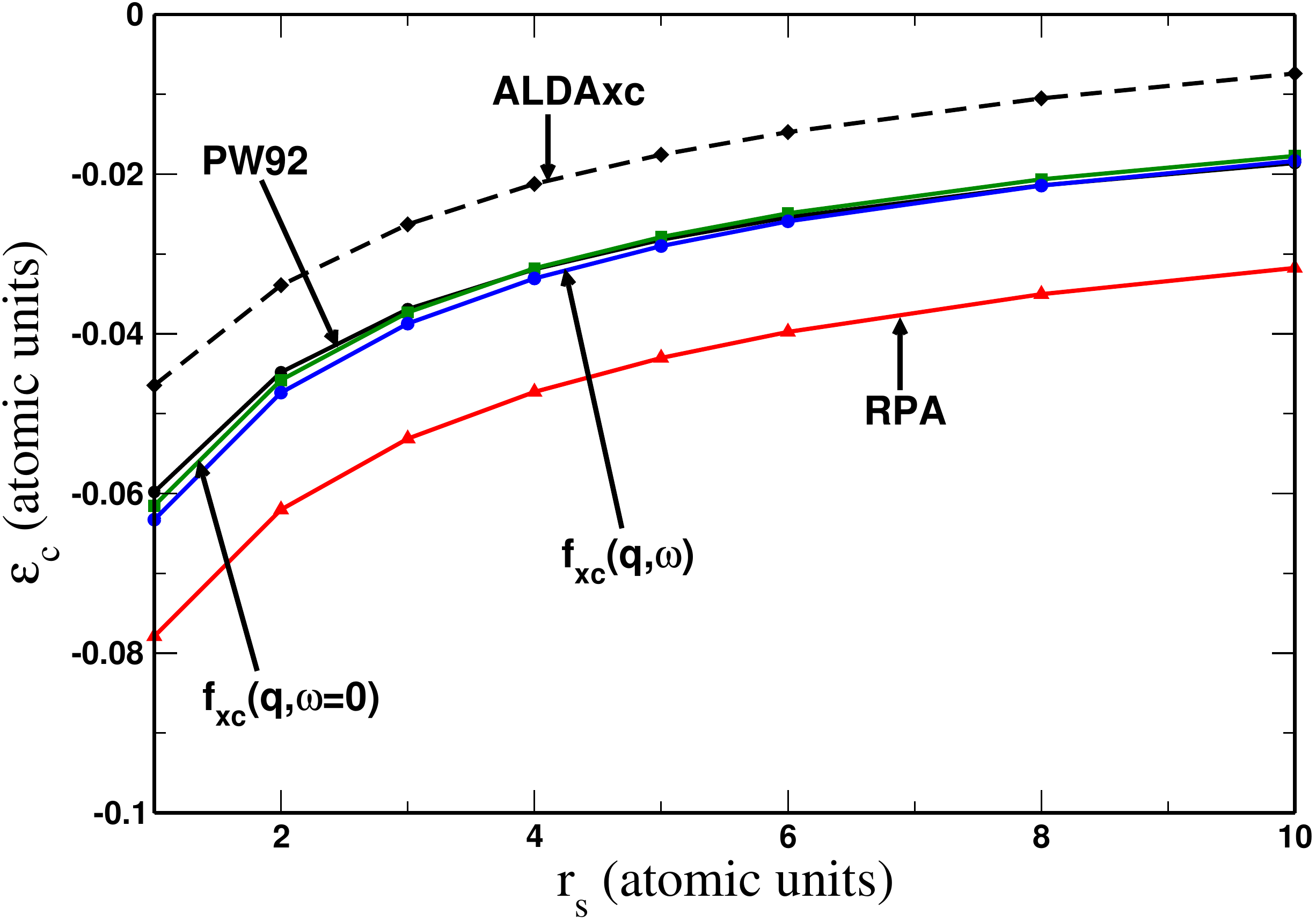}
	\caption{Correlation energy per electron for jellium from various kernels, and the
		essentially-exact Perdew-Wang 1992 (PW92) parametrization, versus density
		parameter r$_s$ . The RPA has no kernel (f$_{xc} = 0$). The adiabatic local density approximation (f$_{xc} =$ f$_{xc}$(0,0)),
		the static MCP07 kernel of Eq. (13) (f$_{xc} =$ f$_{xc}$(q,0)), and the full dynamic kernel of Eq. (24) (f$_{xc} =$ f$_{xc}$(q,$\omega$)) are also tested here. The wavevector dependence and frequency dependence make the kernel f$_{xc}(q,\omega)$ less negative (Figs. 1 and 5), which moves the kernel-corrected correlation energy closer to RPA, in which the kernel is zero.}
\end{figure}

\pagebreak
\newpage
\end{document}